\documentclass[twocolumn,showpacs,amssymb,prd]{revtex4}

\usepackage{graphicx}% Include figure files
\usepackage{dcolumn}% Align table columns on decimal point
\usepackage{bm}% bold math
\usepackage{epsfig}
\begin{document}
\newcommand{\gsim}{\hbox{\rlap{$^>$}$_\sim$}}
\newcommand{\lsim}{\hbox{\rlap{$^<$}$_\sim$}}

%\preprint{APS/123-QED}

\title{On the origin of the diffuse extragalactic gamma-ray background radiation}

\author{Shlomo Dado}
\affiliation{Department of Physics and Space Research
Institute, Technion, Haifa 32000, Israel}

\author{Arnon Dar}
\affiliation{Department of Physics and Space Research
Institute, Technion, Haifa 32000, Israel}

\author{A. De R\'ujula}
\affiliation{Theory Unit, CERN,
1211 Geneva 23, Switzerland;
Physics Department, Boston University, USA}

\date{\today}% It is always \today, today,

\begin{abstract}

We show that inverse Compton scattering of cosmic-microwave-background and 
starlight photons by cosmic-ray electrons in the interstellar and 
intergalactic space explains well the spectrum and intensity of the 
diffuse gamma-ray background radiation (GBR), which was measured by EGRET 
aboard the Compton Gamma Ray Observatory (CGRO) in directions away from 
the Galactic disk and centre. The Gamma Ray Large Area Space Telescope 
(GLAST) will be able to separate the Galactic foreground from
the extragalactic gamma-rays, and to provide stringent tests of the theory.
 
\end{abstract}


\maketitle 

\begin{figure}[]
\centering
\epsfig{file=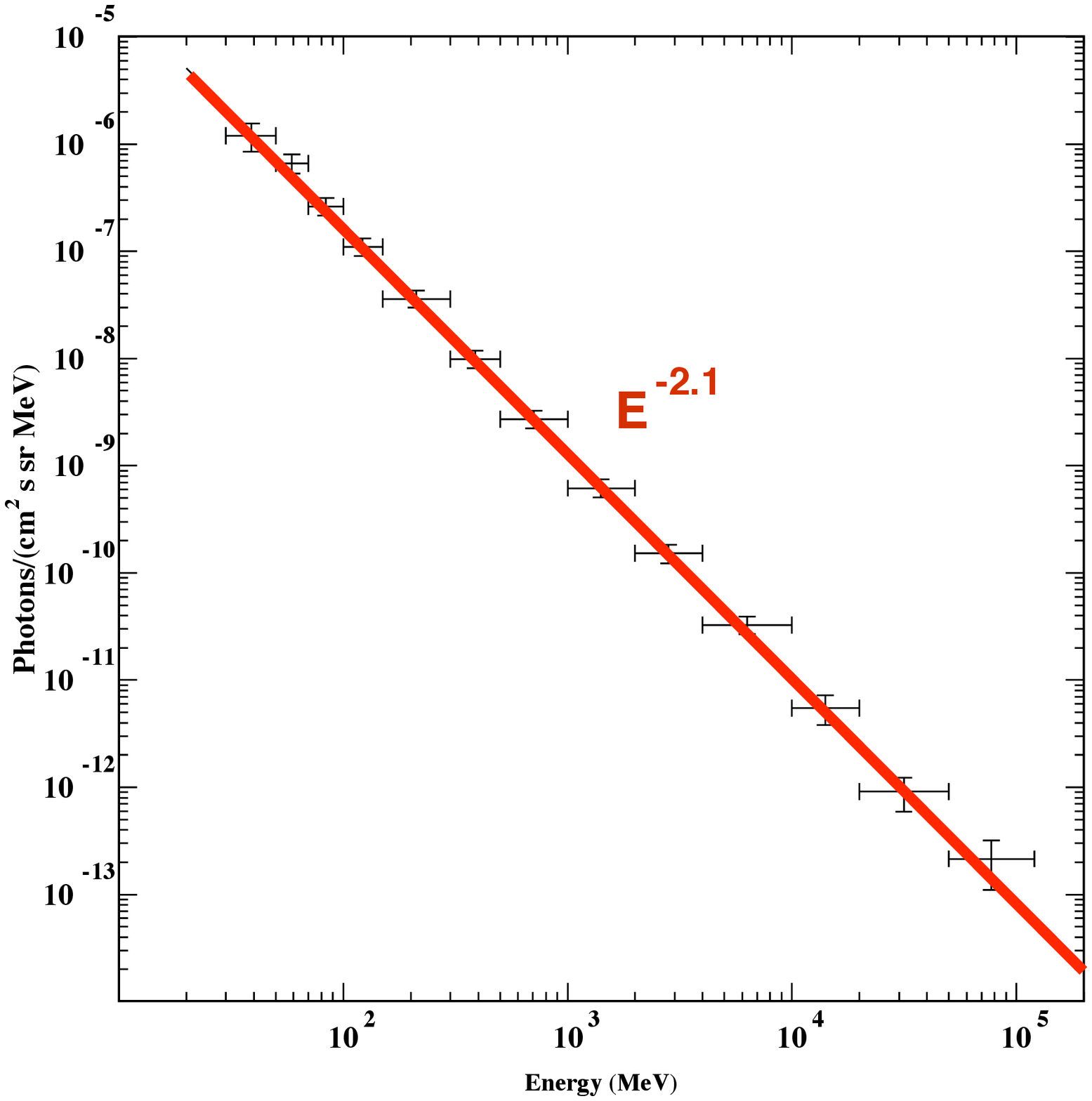,width=9.2cm}
\vspace*{-20pt}
\caption{The GBR spectrum, inferred by EGRET \cite{Sreek}.
 The line is their best power-law fit.}
\label{GBRspectrum} \end{figure}

The intensity and spectrum of the diffuse $\gamma$ radiation  
was measured by  EGRET 
aboard the Compton Gamma Ray Observatory (CGRO).
An extragalactic gamma-ray `background'
radiation (GBR) was inferred ~\cite{Sreek} from  the 
extrapolation of these measurements 
(in directions away from the Galactic disk and center) 
to zero column density, which should eliminate the Galactic 
contributions of bremsstrahlung from cosmic-ray electrons (CREs), and 
$\pi^0$ production by cosmic ray (CR) nuclei. This GBR flux in the 
observed range of 30 
MeV to ${\rm 120~GeV}$, shown in Fig.~\ref{GBRspectrum}, is well 
described by a power-law:
\begin{equation}
\!{dF_\gamma\over dE}\!\simeq\! (2.7\pm 0.1)\times 10^{-3} 
 \left[{E\over {\rm MeV}} 
\right]^{-2.1}\! {1\over {\rm~cm^{2}~s~sr~MeV}}\,.
\label{photons}
\end{equation}

The spectral index of the GBR is the 
same, $2.1\pm 0.03,$ in all sky directions away from the Galactic disk 
\cite{Sreek}. The normalization of the GBR flux in different 
directions was found to be normally distributed around the value 
in Eq.~(\ref{photons}). These results were used to argue for a 
{\it cosmological} (extragalactic) origin of the GBR \cite{Sreek}. A large 
number of putative sources have been proposed, from the quite conventional 
to the decisively speculative. Perhaps the most conservative hypothesis 
is that the GBR is  the sum of $\gamma$-ray 
emissions from unresolved active galactic nuclei (AGNs) \cite{Bignami}. The 
fact that all AGN detected by EGRET  
are blazars with a power-law $\gamma$-ray spectrum with an average index 
$2.15\pm 0.04$, compatible with that of the GBR, supports this hypothesis 
\cite{Chiang}, but later studies have shown that only $\leq 25\%$ of the 
GBR can result from unresolved AGNs \cite{Mukherjee}. Geminga-type 
pulsars, expelled into the Galactic halo by asymmetric supernova 
explosions, could also be abundant enough to explain the GBR \cite{Dixon}. 
Other suggestions include cosmic-ray interactions in galaxy clusters and 
groups \cite{Dar1995}, and fossil radiation from shock-accelerated CRs 
during structure formation \cite{Loeb2000}. More exotic hypotheses are a 
baryon-symmetric Universe \cite{Stecker1971}, now excluded \cite{Cohen}, 
primordial black-hole evaporation \cite{Hawking}, supermassive black holes 
at very high redshift \cite{Gnedin}, and the annihilation of dark-matter 
particles \cite{Silk}.

The EGRET GBR data in directions away from the Galactic disk and centre 
show a significant deviation from isotropy, clearly correlated with the 
structure of the Galaxy and our position relative to its centre 
\cite{DD2001}. This advocates a large Galactic contribution to the GBR.  
In \cite{DD2001} it was shown that the EGRET GBR could be dominated by 
inverse Compton scattering (ICS) of the cosmic microwave background 
radiation (MBR) and 
starlight by Galactic cosmic ray electrons, provided that the 
Galactic cosmic-ray halo is large enough. Indications of a large Galactic 
contribution to the GBR were found by means of a wavelet-based 
``non-parametric'' model-independent approach 
\cite{Dixon}. Other authors \cite{Strong} also found that the contribution 
of inverse Compton scattering of starlight and microwave background 
radiation photons by Galactic CREs is presumably much larger than 
expected.  Earlier evidence that ICS by CREs in the Galactic halo 
contributes significantly to the GBR at large Galactic latitudes was 
reported in \cite{Chen}:  $\sim 30\%$ of the intensity of the GBR at large 
latitudes is correlated to the diffuse Galactic radio emission at 408 MHz, 
which is dominated by synchrotron radiation from the same CREs that 
produce $\sim$ 100 MeV $\gamma$-rays by ICS from the Galactic star-light.

The uniformity of the GBR spectral index over the whole sky, despite a 
large Galactic contribution correlated with the structure of the galaxy's 
halo and our position relative to its centre, suggests similar origins of 
the Galactic and extragalactic contributions. In this letter we argue that 
ICS of the cosmic microwave background radiation  by 
extragalactic CR electrons \cite{DD2006}, when added to the Galactic 
foreground, explains the EGRET GBR. The extragalactic component is 
calculated directly from the CR luminosity of the main putative cosmic 
accelerators \cite{DD2006}: supernova explosions and accreting massive 
black holes in active galactic nuclei (AGN). Unlike in previous estimates of the 
contribution from AGN \cite{Bignami},\cite{Chiang},\cite{Mukherjee}, 
which included only the $\gamma$-ray emission from blazars (AGN with jets and 
$\gamma$ rays beamed in our direction), we calculate the much larger 
contribution from CREs injected into the intergalactic 
medium (IGM)  in arbitrary directions by all AGN jets, and subsequently
isotropized by the IGM magnetic fields. The Galactic component 
was calculated as in \cite{DD2001} from our estimated Galactic CR 
luminosity \cite{DD2006} and from the locally observed flux and spectrum 
of CREs. We show that the observed spectrum, intensity and angular 
dependence of the EGRET GBR are correctly predicted.

The energy spectrum of CREs near Earth \cite{AMS}, with energy 
$E_e>5$ GeV, is well described by:
\begin{equation}
{dF_e\over dE_e}\!=\!(2.5 \pm 0.5)\!\times\! 10^5 \left 
[E_e\over {\rm MeV} \right]^{-3.2\pm 0.1}{1\over {\rm 
cm^2\, s\, sr\, MeV}}. 
\label{CREs}
\end{equation} 
The spectral index is predicted by the Cannonball (CB) model, wherein CRs are 
particles of the interstellar medium accelerated by relativistic 
``cannonballs" ---emitted in core-collapse supernova (SN) explosions--- to 
a ``source" spectrum with a power-law index, $\beta_s=13/6\approx 2.17$ 
\cite{DD2006}.  Energy loss by synchrotron emission in magnetic fields and 
ICS of radiation change $\beta_s$ for CREs to $\beta_e=\beta_s+1\approx 
3.17$. Radio observations of synchrotron radiation emitted by CREs in the 
Galaxy, external galaxies, galaxy clusters and AGN, 
support this predicted universal spectrum of high-energy CREs.

The temperature and mean
energy of the MBR are ${T_0\!=\!2.725}$ K and
$\epsilon_0\!\approx\! 2.7\, k\,T_0\!\approx\! 0.64\,{\rm meV}$
\cite{MandF}. Starlight has 
 $\epsilon_1\!\sim \!1$ eV.
Consider the ICS of these radiations by CREs. 
The mean energy  of the upscattered photons is:
\begin{equation}
{\bar E_\gamma(\epsilon_i) \approx  {4\over 3}\,\left({E_e\over m_e\,
c^2}\right)^2\,\epsilon_i}  .
\label{loss}
\end{equation}
The ICS of the microwave background  and starlight photons by 
CREs produces a GBR with a spectrum 
which is a convolution \cite{Felten} of the CRE  
spectrum with a thermal target spectrum. 
The result can be approximated by:
\begin{equation}
{dF_\gamma\over dE}
\propto{dE_e\over dE}~
\left[ {dF_e\over dE_e}
\right]_{E_e^i}\,, \; {E}_e^i\equiv m_e\, c^2
\sqrt{{3\, \bar E_\gamma\over 4\,\epsilon_i}}\; ,
\label{ICSrad}
\end{equation}
with ${E_e^i}$ obtained from Eq.~(\ref{loss}) by inverting
${\bar E_\gamma}$.
Introducing the electron flux of
Eq.~(\ref{CREs}) into Eq.~(\ref{ICSrad}), we obtain:
\begin{equation}
{dF_\gamma\over dE}\propto E^{-(\beta_e+1)/2}\simeq E^{- 2.08}.
\label{ICSphotpred}
\end{equation}
The predicted index agrees with the measured one, 
${2.10\pm 0.03}$ \cite{Sreek}. 
Given Eq.~(\ref{loss}), CREs of energy $E_{\rm MBR}\!\geq\! 96$ GeV 
produce the GBR above 30 MeV by ICS of the 
current ($z\!=\! 0$) MBR; CREs with energy $E_\star\!\geq\! 2.4$ GeV
suffice for ICS on starlight. Let
$\sigma_{_{\rm T}}\!\approx\! 0.65\times 10^{-24}\, {\rm cm^{-2}}$ be the
Thomson cross-section and let $U_i\!=\!n_i\,\epsilon_i$. In our 
neighbourhood, $U_\star\!\sim\!U_{\rm MBR}\!=\!0.26$ eV cm$^{-3}$.
For electrons of energy $E_i$, the radiation cooling times are 
$\tau_{\rm rad}(i)\! =\! 3\, m_e^2\, c^3/(4\,  \sigma_{_{\rm T}} \, E_i\,  U_i)$,
so that locally $\tau_{\rm rad}(\star)\!\sim\!6\times10^8$ y and
globally $\tau_{\rm rad}({\rm MBR})\!\sim\!1.3\times10^7/(1+z)^4$ y.
These numbers are much shorter than a Hubble time.  
%to a good approximation all electrons relevant to GBR production
%have cooled and radiated.
ICS of MBR photons dominates the production
of the extragalactic GBR, as we argue next.

Adopt a standard cosmology with 
%a Hubble constant
$H\!\approx\! 70$ km s$^{-1}$ Mpc$^{-1}$ and $(\Omega,\, \Omega_M,\,
\Omega_\Lambda) =(1,\,0.27,\,0.73)$, for which the age of the
Universe is  $\sim\!H^{-1}\simeq 14$ Gy.
% Let $U_\gamma$ be the
%energy density of the radiation field, and
%$U_B = B^2 / ( 8 \pi )$ the energy density of a magnetic field $B$.
%Given Eq.~(\ref{loss}), CREs with energy $\geq 96$ GeV are
%needed to produce the GBR above 30 MeV by ICS of the 
%current ($z\!=\! 0$) MBR. 
%The radiative lifetime of these electrons:
%\begin{equation}
%\tau_{\rm rad}(z)  = {3\, m_e^2\, c^3\over
% 4\,  \sigma_{_{\rm T}} \, E_e\,  (U_\gamma+U_B)}<
% {4\times 10^8\over (1+z)^4}\;{\rm y,}
% \label{Radloss1}
%\end{equation}
%is much shorter than the Hubble time (the limit is for $E_e\!>\!3$ GeV, 
%$U_\gamma\!=\!U_{\rm MBR}$). 
For a  Galactic magnetic field $B\!\sim\!3\,\mu$G,
$U_B\!\sim\!U_{\rm MBR}$; synchotron cooling and emission
are locally relevant \cite{DD2001}. 
%The solar, star-light and MBR contributions 
%to the spectrum are comparable for the Galactic component of the 
%GBR \cite{DD2001}. 
In our model, CBs transfer their kinetic
energy to CRs all along their trajectories, which extend from the SN-rich
inner galaxies to their halos and beyond. In galactic halos and  
galaxy clusters, $B\!<\!3\, \mu$G, and in the IGM, $B\sim 50\,$nG 
\cite{DD2005}. In both places starlight is irrelevant, and
ICS of the MBR whose energy density increases with $z$ like $(1+z)^4$
dominates over synchrotron losses on the 
magnetic field. Thus, we calculate the intensity 
of the extragalactic GBR from the conclusion that 
%practically all 
the kinetic energy of CREs in the Universe 
with a lifetime shorter than the Hubble time 
has been converted by ICS of the MBR to 
$\gamma$-rays with the predicted spectrum of 
Eq.~(\ref{ICSphotpred}).

In the CB model, the  main accelerators of high-energy CREs are the
relativistic jets of supernovae (SNe)  and AGNs \cite{DD2006}.  
%Consider SNe first. 
The SN
rate is proportional to the star-formation rate $R_{_{\rm
SN}}(z)\!\propto\! R_{_{\rm SF}}(z)$, with $R_{_{\rm SN}}(0)\approx
10^{-4}\, {\rm Mpc^{-3}\,yr^{-1}}$ \cite{Capellaro}. The observations are
well represented by $R_{_{\rm SF}}(z)/R_{_{\rm SF}}(0)\!\approx\! (1+z)^4$
for $z\!<\!1$ and $R_{_{\rm SF}}(z)\!=\!R_{_{\rm SF}}(1)$ for $z\geq
1.2$ \cite{SFR}. Let $E_k\!\approx\! 2\times 10^{51}$ erg be the mean
energy release in CRs per SN \cite{DD2006} and let $f_e$ be the fraction
of the luminosity in CREs out of the total  $L_{_{\rm CR}}$ in
CRs. The CB model does not predict $f_e$, we  assume that it is equal
to the ratio of the Milky Way's luminosity in CREs to its total luminosity
in CRs:
\begin{equation}
\!f_e\approx \!\!{\int
{dF_e[{\rm MW}]\over dE}\, {E\,dE\over \tau_{\rm 
rad}}
\Big/\!\!\int {dF_{_{\rm CR}}[{\rm MW}]\over dE}\, {E\,dE\over \tau_{\rm esc}}}
\approx {1\over 40}
\label{fe}
\end{equation}
where $\tau_{\rm esc}\approx
2\times 10^8\ (E/{\rm GeV})^{-0.6}$ yr
is the mean escape time of CR protons and electrons from the Galaxy
and its halo by diffusion in its magnetic field \cite{DD2006}. In the CB
model the volumes occupied by electron and proton CRs are similar, 
because CBs generate CRs all along their trajectories, which constitute a
dense mesh in the Galaxy and its halo.
%at any point in time, 
%and extend to its halo and beyond.
%(Even close-by CBs are very hard to observe,
%because of the extremely forward nature of their radiation). 
The integrals in Eq.~(\ref{fe}) extend from a lower fixed Lorentz factor,
which drops from the ratio $f_e$, since the integrands are source spectra,
identical in the CB model for electrons and protons.

The energy of CRE
made by SN jets is converted, 
above $E_e\!\sim\!100$ MeV, to photon energy.
Their contribution to the GBR spectrum satisfies:
\begin{equation}
\int_{E_{c}}\! \!{dF_\gamma\over dE}\,E\,dE\approx\!
{c\, L_e/R_{_{\rm SF}}(0)\over 4\pi\,H_0}\!\!\int\!\!
{dz\,R_{_{\rm SF}}(z)/(1+z)^{\beta_s}\over 
\sqrt{\Omega_M\, (1+z)^3+\Omega_\Lambda}},
\label{SNcont}
\end{equation}
where $L_e\!=\!f_e L_{_{\rm CR}}\gamma_e^{-1/6}\! \approx\!
f_e R_{_{\rm SN}}E_k\gamma_e^{-1/6}$ is
the mean luminosity density of CREs with Lorentz factor above
$\gamma_c=\sqrt{3\, E_c/4\,\epsilon_0}$.
Hence, Eq.~(\ref{SNcont}) yields for the contribution to the
GBR from extragalactic SNe:
\begin{equation}
 {dF_\gamma\over dE}\simeq 0.9\times 10^{-3}
 \left[{E\over {\rm MeV}} \right]^{-2.08}
{1\over \rm cm^{2}~s~sr~MeV}\; .
\label{SNgam}
\end{equation}

%The dependence of the energy integrals in Eqs.~(\ref{fe}), and (\ref{SNcont})
%on their low energy cutoffs is weak: $E_{c}^{-0.17}$ and  $E_{c}^{-0.085}$,
%we chose $E_{c}=???$ to obtain Eq.~(\ref{SNgam}).

Powered by mass accretion onto massive 
black holes, AGNs eject powerful relativistic jets whose kinetic energy is 
transferred mainly to CRs. The kinetic power of these jets has 
been estimated from 
their radio lobes, assuming equipartition between CR- and 
magnetic field energies 
and an energy ratio $f_e$ similar to that observed in our 
Galaxy. It was estimated \cite{Kronberg} that
AGNs with a central black hole of  $M\!\simeq\! 10^8\, M_\odot$
inject $\approx\! 10^{61-62}$ erg into the intergalactic space, 
mostly during their $\sim\! 10^8$ y  bright phase around redshift 
$z\!=\!2.5$. In search for an upper bound, we assume that the kinetic 
energy release 
in relativistic jets is the maximal energy release from mass accretion 
onto a Kerr black hole ($\approx\! 42\%$ of its mass), and that 
this energy is equipartitioned between magnetic 
fields and cosmic rays \cite{DD2006} with a fraction
$f_e$  of the CR energy carried by  electrons.
These CREs also cool rapidly by ICS of the MBR. 
The energy of CREs 
whose radiative cooling rate $\tau_{\rm rad}(z)$ is larger than the cosmic 
expansion rate, $H(z)$, is converted 
to $\gamma$-rays. Their energy is redshifted by $1+z$ by the cosmic 
expansion. Using a black hole density, 
$\rho_{_{\rm BH}}(z=0)\!\sim\!2\times 10^5\, M_\odot\,
{\rm Mpc^{-3}}$ 
in the current Universe \cite{Tremaine},
and the CB-model 
injection spectral index \cite{DD2006},
our estimated contribution from AGNs to the extragalactic GBR flux,
\begin{equation}
{dF_\gamma\over dE}\simeq 
{2.4\times 10^{-3}\, c\, f_e\, \rho_{_{\rm BH}}\, c^2\over 4\,\pi\, 
(1+z)\, {\rm MeV^2}} 
\left[{E\over {\rm MeV}} \right]^{-2.08}\,,
\label{AGNs1gam}
\end{equation} 
is,
\begin{equation}
{dF_\gamma\over dE}\simeq 
4.0\times 10^{-4} 
\left[{E\over {\rm MeV}} \right]^{-2.08}~
{1\over \rm cm^{2}~s~sr~MeV}\, .
\label{AGNs2gam}
\end{equation} 
This upper limit is smaller than the SN result of Eq.~(\ref{SNgam}).

%Because of Feynman scaling, the GBR from $\pi^0$ production and decay 
%in hadronic CR collisions in the ISM and IGM has the same 
%power-law index as that of CRs \cite{Dar1995}, i.e.~$-2.77$ in the ISM of 
%galaxies and $-2.17$ in the IGM inside and outside galaxy clusters 
%\cite{DD2006}. However, this contribution to the extragalactic GBR 
%is much smaller than that of CREs and need not be discussed here.

\begin{figure}
\begin{center}
%\vspace*{-1.6cm}
%\hspace*{-1cm}
\epsfig{file=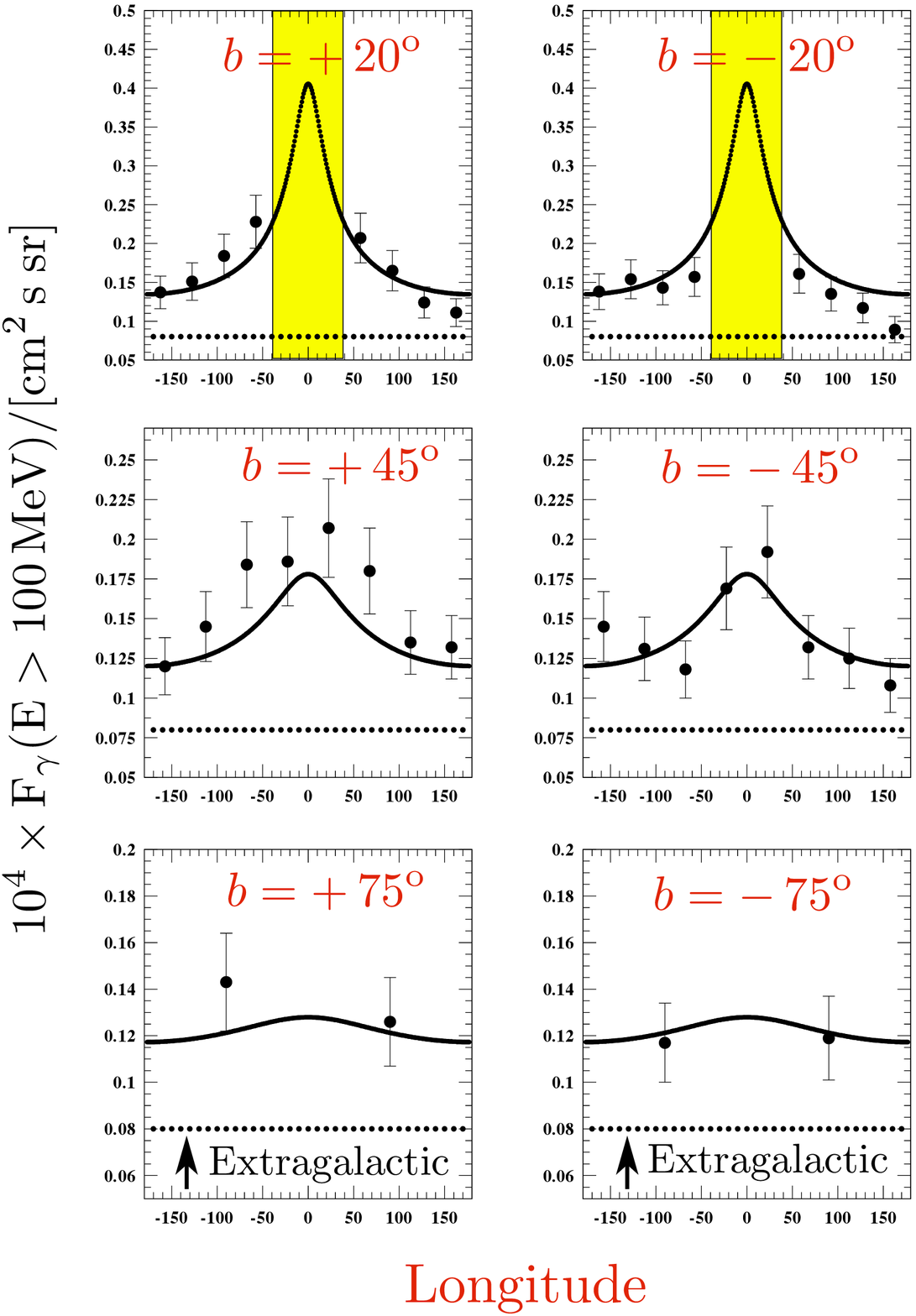,width=9cm}
%\vspace*{-14.6cm}
\caption{The flux of GBR photons above 100 MeV: comparison
between EGRET data and our model
for ${h_e\!=\!8}$ kpc, ${\rm \rho_e\!=35\!}$ kpc, as functions
of longitude $l$ at  fixed latitudes $b$. The shaded domain
is EGRET's mask. 
%The AGN contribution is subdominant:
%compare Eqs.~(\ref{SNgam}) and (\ref{AGNs2gam}).
Notice that the vertical scales do not start at zero. }
\vspace*{-0.5cm}
\end{center}
\end{figure}

The GBR contains a considerable Galactic foreground due to ICS of MBR, 
starlight and sunlight photons by Galactic CREs \cite{DD2006}. The 
convolution of a CRE power-law spectrum with a photon thermal 
distribution \cite{Felten} can be approximated very simply \cite{DD2001}.
Using the index $i$ to label the MBR, starlight and sunlight fluxes, we 
have:
\begin{equation}
{{dF_\gamma\over dE_\gamma}
\simeq {N_i(b,l)~\sigma_{_{\rm T}}}~{dE^i_e\over dE_\gamma}~
\left[ {dF_e\over dE_e}
\right]_{E_e={E}_e^i}}\, ,
\label{ICSGal}
\end{equation}
where $N_i(b,l)$ is the column density of the radiation field 
weighted by the distribution of CREs in the direction $(b,\, l)$, 
and ${ E_e^i}$ is given in  Eq.~(\ref{ICSrad}). The distribution of the
non-solar starlight is approximated as $\propto 1/r^2$, with $r$ the
distance to the Galactic centre, and the CREs are assumed to be
distributed as a Gaussian  ``CR halo" \cite{DD2001}.
Naturally, the results depend crucially on the size and shape
of this halo. In this note we use our 
updated estimate of the CR halo 
\cite{DD2006}: a Gaussian distribution with a scale length of
$\rho_e\!=\! 35$ kpc in 
the Galactic disk, as we used in \cite{DD2001}, but a scale height of
$h_e\!=\!8$ kpc perpendicular to the disk \cite{DD2006}
instead of the $h_e\!=\!20$ kpc used in \cite{DD2001}. 
The justification for this change is as follows:
The radio emission of ``edge-on" galaxies  --interpreted as synchrotron 
radiation by electrons on their magnetic fields-- offers direct 
observational evidence for CREs well above galactic disks 
(e.g.~\cite{Duric1998}). For the particularly well observed case of NGC 
5755, the exponential scale height of the synchrotron radiation is ${\cal{O}} 
(4)$ kpc. If the CRs and the magnetic field energy are in equipartition, 
they should have similar distributions, and the Gaussian scale height 
${ h_e}$ of the electrons ought to be roughly twice that of the 
synchrotron intensity, which reflects the convolution of the electron- and 
magnetic-field distributions. The inferred value is ${h_e}\sim 8$ kpc.
The corresponding volume of the Galactic CR halo is 
$V_{_{\rm CR}}\!=\!(\pi)^{3/2}\, 
\rho_e^2\, h_e \!=\! 1.6\times 10^{69}\, {\rm cm^3}$.
The SN rate in the Galaxy is $R_{_{\rm SN}}[\rm MW]\!\sim$ 2 per century, and
its predicted total luminosity in CRs  is
$L_{_{\rm CR}}\!\approx\! E_k\, R_{_{\rm SN}}[\rm MW]\approx 4\times 10^{49}\, 
{\rm erg\, y^{-1}}$. 
The CR confinement volume must obey the constraint:
\begin{equation}
L_{_{\rm CR}}\sim V_{_{\rm CR}}\times
{4\pi \over c}\, \int {dE\over \tau_{\rm esc}}\,E\,{dF_p\over dE}\; .
\label{CRlum2}
\end{equation}
Our estimated $\tau_{\rm esc}$
and the observed (or fitted) spectrum of CRs
\cite{AMS,Haino} yield the expected
$V_{_{\rm CR}}\!\approx\! 1.6 \times 10^{69}$  cm$^3$.
The volume 
% $V_{_{\rm CR}}\!\approx\! 6.6\!\times\! 10^{68}$ cm$^3$,
inferred from a leaky-box model fit to the Galactic GBR \cite{SM} 
is smaller by a factor $\approx 2.5$ than our estimate, reflecting the
shorter confinement time of CRs estimated in leaky-box
models from the abundance of unstable CRs \cite{Connell}, 
and the higher contribution  assumed in \cite{SM} for the 
extragalactic GBR.

In Fig.~2 we compare the observed GBR with our predictions, as functions 
of Galactic coordinates. The prediction is a sum of a $(b,\, l)$-dependent 
Galactic foreground produced by ICS of the MBR, starlight and sunlight, 
and a uniform extragalactic GBR. The result has $\chi^2/{\rm dof}\!=\!0.85$, a 
vast improvement over the constant GBR fit by EGRET, for which 
$\chi^2/{\rm dof}\!=\!2.6$. The ratios of $l$-integrated extragalactic to galactic fluxes 
are $\sim\!0.5$, 0.9, 1.5, for $\vert b\vert=20^o$, $45^o$, $75^o$. 
The `foreground' component of the $\gamma$ `background' 
is $\sim\!50\%$ of the total radiation.

We conclude that the GBR can be explained by 
standard physics, namely, ICS of MBR and starlight by CREs from the two 
main CR sources in the universe: SNe and AGNs. At $E_\gamma\! >\! 100$ 
GeV, most of the extragalactic GBR is absorbed by pair production 
on the CMB \cite{Salamon} and the diffuse 
GBR reduces to the Galactic foreground. This suppression, and a
decisive determination of the angular dependence in Fig.~2, should be 
observable by GLAST. 
%The $\gamma$ radiation from the halo of Andromeda
%may also be observable .

\end{document}